\documentclass[12pt]{article}

\def\half{{\scriptstyle\frac{1}{2}}}
\def\r{\hat\rho}
\def\P{\hat P}
\def\G{\Gamma}
\def\L{{\cal L}}
\def\s{\sigma}
\def\p{\hat p}\def\x{\hat x}
\def\a{\alpha}
\def\xt{\tilde x}\def\pt{\tilde p}\def\Gt{\tilde\G}

\def\C{{\bf C}}

\def\ket#1{|#1\rangle}
\def\bra#1{\langle#1|}

\def\be{\begin{equation}}
\def\ee{\end{equation}}
\def\bea{\begin{eqnarray}}
\def\eea{\end{eqnarray}}

\newcommand{\lb}{\label}
\newcommand{\I}{\mbox{i}}
\newcommand{\D}{\mbox{d}}
\newcommand{\E}{\mbox{e}}

\begin{document}
\begin{titlepage}

\vskip 1cm
\begin{center}

{\large\bf EXACT POSITIVITY OF THE WIGNER AND P-FUNCTIONS
           OF A MARKOVIAN OPEN SYSTEM}
\vskip 1cm
{\bf Lajos Di\'{o}si}
\vskip 0.4cm

Research Institute for Particle and Nuclear Physics,\\
P.O. Box 49, H-1525 Budapest 114, Hungary.

\vskip 0.7cm
{\bf Claus Kiefer}
\vskip 0.4cm
Institute for Theoretical Physics, University of Cologne,\\
Z\"ulpicher Str.~77, 50937 K\"oln, Germany.

\end{center}
\vskip1cm
\small
\begin{center}
{\bf Abstract}
\end{center}
\begin{quote}
We discuss the case of a Markovian master equation for an open system,
as it is frequently found from environmental decoherence.
We prove two theorems for the evolution of the quantum state.
The first one states that for a generic initial state the corresponding 
Wigner function becomes strictly positive after a finite time has elapsed.
The second one states that also the P-function becomes exactly positive
after a decoherence time of the same order. Therefore the density matrix 
becomes exactly decomposable into a mixture of Gaussian pointer states.
\end{quote}
\normalsize
\date{\today}
\end{titlepage}

\section{Introduction}

The study of Markovian open systems is of general interest.
One of the reasons is that the coupling of open 
systems to their ubiquitous environment often leads to master
equations which are local in time \cite{deco}. The interaction of
dust particles with air molecules or radiation, for example, delocalises
any interference terms into correlations with the environmental
degrees of freedom
on a short decoherence timescale. Thereafter, the dust particle
can be perfectly described by a Markovian master equation for its
density matrix $\r(t)$: 
\be
\frac{d\r}{dt}\equiv\L\r
                 =-\frac{\I}{2m}[\p^2,\r]-
               \frac{D}{2}[\x,[\x,\r]]\ .
 \lb{master}
\ee
Such an equation results frequently from an interaction with
the environment in cases where friction is negligible \cite{deco}.
The strength of the coupling is given by the parameter $D$.
The first term in (\ref{master}) would lead to unitary spreading,
whereas the second term would lead to nonunitary localisation.
For a wave function of characteristic width $\sigma$, these effects
act on timescales $m\sigma^2$ and $1/D\sigma^2$,
respectively. 
Both effects are thus balanced for an ``equilibrium width'',
given approximately by
\be
\sigma_0 = \left(Dm\right)^{-1/4}\ , \lb{width}
\ee 
see \cite{Dio87a,deco}.
The corresponding timescale is 
\be
t_0=\sqrt{m/D} 
\lb{time}
\ee
and will set the timescale for decoherence.

The oldest way to elucidate a quantum state in terms of a
pseudo-classical
distribution of phase-space variables $\G\equiv(x,p)$ is due to Wigner. 
Instead of the density matrix, one discusses the Wigner function
\be
W(\G)\equiv\int\bra{x-\half r}\r\ket{x+\half r}\E^{ipr}\D r
\lb{Wigner}
\ee
in order to study aspects of an open system \cite{deco}. 
Normalization holds with the choice $d\G=dpdx/2\pi$ of the phase-space
volume element. The Wigner function $W(\G;t)$ corresponding to 
(\ref{master}) satisfies the Fokker-Planck equation
\begin{equation}
\frac{dW}{dt} = -\frac{p}{m}\frac{\partial W}{\partial x} 
                 +\frac{D}{2}\frac{\partial^2 W}{\partial p^2}\ .
\lb{FPW}
\end{equation}
In a general situation, the Wigner function is negative in some
regions of phase space. For this reason it cannot be regarded as
a probability distribution. If decoherence occurs, the general expectation
is that these negative parts are smoothed out in the course of time.
Many examples support this expectation. In Sect.~2 of this paper
we shall prove a much stronger statement: After a
certain decoherence time, the Wigner function becomes
{\em strictly positive}. This is very different from the behaviour
of the density matrix whose nondiagonal terms (describing
interferences) become zero only asymptotically, i.e., they
remain nonzero at any finite time. 

Through the process (\ref{master}) or, equivalently, (\ref{FPW}) 
the position basis is distinguished as the preferred basis with 
respect to which no interferences can be observed and which remains 
a robust basis in time.
Such a classical basis is usually called the pointer basis.
It is the basis which remains most stable against the influence
of an environment. How can the pointer basis be determined?
We have shown in \cite{DK} that this can be achieved by three
different methods which all lead to the same results.
The first method invokes the principle of ``Hilbert-Schmidt
robustness'' stating that the pointer states mimic the local
nonunitary evolution as closely as possible with respect to
the Hilbert-Schmidt norm. A unique set of Gaussian pointer states 
has thereby been found. The second method demands that the local production
of entropy be minimal (``predictability sieve'').
Again, this has led to a set of unique
Gaussian pointer states with practically the same width as the
previous ones. The last method invokes quantum state diffusion and
leads again to Gaussian pointer states with the same width.

In Sect.~3 we introduce the overcomplete set of Gaussian pointer
states $\ket{\G}$. Then we define additional phase-space distributions: 
The (generalized) Q- and P-functions are related to the Wigner
function by Gaussian coarse-grainings.
In the discussion of the local entropy production \cite{DK} 
we have already used
the important fact that the density matrix can be decomposed 
exactly into a mixture of Gaussian states after a finite decoherence
time. In Sect.~4 of the present paper we shall prove this 
theorem -- the positivity of the P-function --
which, in fact, holds for arbitrary initial states. The technical
details are relegated to Appendices~A, B, and C.

\section{Strict positivity of the Wigner function}

We shall now consider the Wigner function, $W(\G;t)$, of our open
quantum system, obeying Eq.~(\ref{FPW}).
We shall prove the theorem that
for {\em any} initial state, $W$ will become positive after a finite
time $t_D$, i.e.,
\be
W(\G;t)\geq 0, \quad t\geq t_D\ .\lb{theoremW}
\ee
$W$ can of course be positive even earlier; the theorem states that
it cannot be negative later. The Wigner function corresponding for a
Gaussian wave function, for example, stays positive for all times.
In fact, as shown e.g. in \cite{Tat83}, Gaussian states are the only
pure states that lead to a positive Wigner function.
It must, however, be emphasised that positivity is only a necessary
requirement for classicality, not a sufficient one. Squeezed
Gaussian states of harmonic oscillators, for example,
 are genuine nonclassical states,
but correspond to a positive Wigner function. 

Since Fokker-Planck equations like (\ref{FPW}) preserve the 
positivity of distribution functions,
one might guess that the set of positive solutions attracts the
indefinite ones, so that any indefinite distribution becomes
positive in the course of time. This is, however, not true. 
Eq.~(\ref{FPW}) is a linear equation. Since it also allows
solutions that remain negative, the superposition principle
will allow indefinite solutions at all times. 

The proof, therefore, has to rely on specific properties of the
Wigner function. It is know that $W$ possesses, in fact, very special
features, see e.g. \cite{Leo}. For example, it obeys
$|W|\leq 1/\pi$. Moreover, its negative parts are always restricted
to small regions in phase space. 

To be concrete, in our proof we shall make use of two properties 
for the Wigner function. The first one exploits its connection
with the {\em Q-function} (see e.g. \cite{Leo}). 
Introducing the normalized Gaussian
\begin{equation}
  g(\G;C)= \frac{1}{C}\exp
\left[-\frac{x^2+p^2}{2C}\right]\ , \lb{gauss2}
\end{equation} 
its convolution with the Wigner function,
$g\star W$, is always 
positive provided $C\geq1/2$ (and may be indefinite otherwise). 
In the marginal case $C=1/2$ this convolution gives the
Q-function which yields the probability distribution for finding
the coherent states in the density operator $\r$, and is therefore
manifestly positive \cite{Leo}. This is why the negative regions
for $W$ are so restricted: the convolution of $W$
with the Gaussian $g(\G;1/2)$ would not yield a positive function
if $W$ contained negative regions that are spread out over regions
with areas much bigger than about $1/2$. Since the case $C>1/2$ represents
a stronger coarse-graining of $W$ than $C=1/2$, it is clear 
that this positivity remains true.

The second property of the Wigner function, which will serve
as an ingredient in our proof, is its invariance under
{\em linear} canonical transformations.
To be precise, if we make such a transformation for both the
quantum operators and the classical canonical variables, one has 
\begin{equation}
(\x,\p)\rightarrow(\x',\p')~~\Rightarrow~~
W(x,p)\rightarrow W'(x',p')=W(x,p)\ .
\end{equation}
This feature is not obvious from the usual expressions defining $W$.
There exists, however, an alternative equivalent form
given by \cite{Dio96}:
\begin{equation} 
W(x,p)=\mbox{tr}\left[\r\{\delta(x-\x)\delta(p-\p)\}_{sym}\right]\ ,
\end{equation}
where we refer to a simple symmetrisation process regarding
the order of $\x$ and $\p$: 
\begin{equation}
\{\hat\Gamma\hat O\}_{sym}\equiv
\frac{1}{2}(\hat\Gamma\hat O+\hat O\hat\Gamma),~~~
\mbox{for}~\hat\Gamma=(\x,\p)~\mbox{and for all}~\hat O.
\end{equation}
This ordering is explicitely invariant for linear
canonical transformations. The invariance of the
Wigner function follows immediately. Note, however, that the
coarse-graining is {\em not} invariant and coarse-grained
Wigner functions, Q-functions in particular, will be
non-invariant even for linear canonical transformations. 
In the simple case of the transformation
$\x'=a\x, \p'=a^{-1}\p$, and assuming a pure state with wave function
$\psi(x)$, the invariance can be seen directly from the standard
integral expression.
Using
\be
\psi'(x')=\frac{1}{\sqrt{a}}\psi\left(\frac{x'}{a}\right)\ ,
\ee
one has 
\be
W'(x',p')=\frac{1}{\pi}\int\ \D y'\exp(2ip'y')\psi^{*'}(x'+y')\psi'(x'-y')\ ,
\ee
where $W'$ denotes the Wigner function with respect to the same state
in the transformed basis, and therefore
\be
W'(x',p')=W(x,p)\ . \lb{invW}
\ee
Using this invariance, one can extend the previous coarse graining
(\ref{gauss2}) to general Gaussians with correlation matrix $\C$: 
\be
g(\G;\C) = \vert\C\vert^{-\half}
          \exp\left[-\G^T\frac{1}{2\C}\G\right] \lb{gengauss}
\ee
(When vectors, $\G$ stands for column and $\G^T$ for row vectors,
respectively.)
Application of a linear canonical transformation
to the convolution $g\star W$ rendering $\C=\sqrt{\vert\C\vert}{\bf I}$
then demonstrates that the sufficient and necessary condition for the
positivity of the coarse-grained Wigner function reads
\be
g(\G;\C)\star W(\G) \geq0\ \ \ \ iff\ \ \vert\C\vert\geq1/4.
\lb{Wpos}
\ee

With this lemma, the proof of the theorem (\ref{theoremW}) becomes
straightforward. The Fokker-Planck equation (\ref{FPW}) imposes 
a progressive Gaussian coarse-graining (\ref{gengauss}) on the initial 
Wigner function: 
\begin{equation}
W(\G;t)=g(\G;\C_W(t))\star W(x-pt/m,p;0) \lb{Wsol}\ ,
\end{equation}
where the time-dependent correlation matrix of the coarse-graining is 
\begin{equation}
\C_W(t)=
Dt\left(\begin{array}{cc}t^2/3m^2&t/2m\\t/2m&1\end{array}\right)
, \lb{BW}
\end{equation}
as can be found from Eqs.~(\ref{FPW},\ref{gengauss},\ref{Wsol}).
The determinant yields
\begin{equation}
\vert\C_W(t)\vert=\frac{D^2t^4}{12m^2}\ .
\end{equation}
It follows from the condition (\ref{Wpos}) that 
the Wigner function is indeed positive for
\be
 \frac{t}{t_0} \geq 3^{1/4}\approx 1.32\ ,
\ee
which is of the same order as the decoherence timescale of
Sect.~1. This completes our proof.

We would like to mention that,
based on more complicated mathematics \cite{KhaTsi87}, the lemma
(\ref{Wpos}) has been stated earlier \cite{KhaTsi92}.
Similarly to the spirit of \cite{Dio87b}
and the present work, it was also suggested in
\cite{KhaTsi87,KhaTsi92} 
that particular quantum features of open systems would disappear
after an estimated finite time.

 
\section{Pointer states, Q- and P-functions}

We are going to consider the overcomplete set of normalized pure 
Gaussian pointer states $\ket{\G}$ obeying
\be
\int\ket{\G}\bra{\G}\D\G=\hat I \ \ .
\lb{compl}
\ee
In Appendix~A we discuss the position representation of these states.
The overlap of two pointer states is a Gaussian:
\be
\vert\bra{\G}\G^\prime\rangle\vert^2
=\exp\left[-(\G-\G^\prime)^T\frac{1}{4\C_{1/4}}(\G-\G^\prime)\right]
\ . \lb{overl}
\ee
The matrix $\C_{1/4}$ is positive and has determinant $1/4$. This 
value for the determinant
 makes the above Gaussian normalized and assures the consistency
of the completeness relation (\ref{compl}) with the 
normalization of the pointer states.
 In fact, $\C_{1/4}$ can be directly identified as   
the matrix of quantum uncertainties of the canonical pair of operators
in the Gaussian pointer states, cf. (\ref{C1}) in Appendix~A.
 We can calculate  $\C_{1/4}$ for simplicity
in the fiducial state $\ket{\G=(0,0)}\equiv\ket{0}$: 
\be\lb{C}
\C_{1/4}\equiv\bra{0}\hat\G\hat\G^T+\mbox{h.c.}\ket{0}\ .
\ee

Given the above overcomplete set of Gaussian states, one can 
introduce the generalized Q- and P-functions, related to the
density operator respectively by
\be
Q(\G)=\bra{\G}\r\ket{\G}
\lb{Q}
\ee
and 
\be
\r=\int P(\G)\ket{\G}\bra{\G}\D\G \ .
\lb{P}
\ee
The Q-function is the probability distribution of the value $\G$.
In a generalized quantum measurement it can be inferred from the 
{\em positive operator valued measure} (see e.g. \cite{Per})
formed by $\ket{\G}\bra{\G}\D\G$. The P-function has
a different meaning. When we expand the density matrix as a sum of
the pointer states, the weighting function is called the P-function.
While the Q-function is non-negative by construction, the P-function
may be indefinite (even ill-defined)
 for generic states. The Q- and P-functions are related 
to the Wigner function by the same Gaussian coarse-graining but in opposite
senses \cite{Leo}:
\be
W(\G)=g(\G;\C_{1/4})*P(\G)\ ,
\lb{WgP}
\ee
\be
Q(\G)=g(\G;\C_{1/4})*W(\G)\ .
\lb{QgW}
\ee
The correlation matrix $\C_{1/4}$
of coarse-grainings is the one that appeared
earlier as the matrix of canonical quantum uncertainties in the
Gaussian pointer states, see (\ref{C}).
All sets of Gaussian pure pointer states are classified in Appendix A.
Details of derivations for (\ref{WgP}),(\ref{QgW}) are given in Appendix B.

The Q-function satisfies a Fokker-Planck equation which is,
according to (\ref{QgW}), the coarse-grained version of the 
Fokker-Planck equation (\ref{FPW})
for the Wigner function. For the evolution 
equation of the P-function we are going to present a more direct 
derivation in Appendix C.


\section{Strict positivity of the P-function} 

In Sect.~3 we have expanded the density operator as a weighted sum of 
Gaussian pointer states, see (\ref{P}). The weight function is called
P-function and it is indefinite for a generic quantum state $\r(0)$.  
We shall now prove that,
due to the open system dynamics (\ref{master}), the P-function becomes 
{\em exactly} positive, 
\be
P(\G;t)\geq0,~~~~t\geq t_D'\ .
\lb{theoremP}
\ee
The density operator $\r(t)$ can thus be 
decomposed {\em exactly} into a statistical mixture of
Gaussian pointer states $\ket{\G}$ (see (\ref{gauss}) below) after
a {\em finite} decoherence time $t_D'$ has elapsed.
We already emphasise here that this holds
for a {\em generic} $\r(0)$
(not necessarily Gaussian). 
This theorem generalises the corresponding statement
made in \cite{Dio87b} for a single choice of Gaussian pointer states
(i.e. of $\C_{1/4}$)
as well as the asymptotic statements proved in \cite{HalZou95,HalZou97}. 

The proof will be reduced to the lemma (\ref{Wpos}) used in the
proof of the positivity of the Wigner function, see (\ref{theoremW})
in Sect.~2. Consider the solution (\ref{Wsol}) for the Wigner
function and substitute the expression~(\ref{WgP}) 
into its left-hand side.
As for the right-hand side, assume that enough time has elapsed so that 
$\C_W(t)-\C_{1/4}$
is a non-negative matrix. Then the convolution factorizes as
\be
g(\G;\C_W(t))=g(\G;\C_{1/4})*g(\G;\C_W(t)-\C_{1/4})
\ee
and can be substituted into (\ref{Wsol}). We obtain
\be
 g(\G;\C_{1/4})\star P(\G;t)
=g(\G;\C_{1/4})\star g(\G;\C_W(t)-\C_{1/4})\star W(x-pt/m,p;0)\ .
\ee
The identical convolutions on both sides cancel each other and 
leave us with the explicit solution for the P-function as the 
coarse-grained Wigner function.
The lemma (\ref{Wpos}) tells us that a generic initial P-function 
becomes non-negative after a time $t$, 
\be
\lb{Ppos}
P(\G;t)\geq0\ \ \ \ \ iff\ \ \vert\C_W(t)-\C_{1/4}\vert\geq1/4 \ .
\ee
Calculating the determinant from (\ref{BW}) and (\ref{C2}) below, 
this condition leads to a cubic equation for $t$, yielding
the numeric estimate
\be
\frac{t}{t_0}\geq1.97\ \ .
\ee
As expected,
therefore, the decoherence timescale $t_D'$ coincides
approximately with $t_0$. This completes the proof. 

One might wonder how much of our results do depend on our particular
Markovian master equation (\ref{master}). 
We expect that our proofs of positivity can be extended
to all master equations of this kind which are at most
quadratic in position and momentum, e.g. for the damped
harmonic oscillator. 
As far as corrections to the Markovian
approximation are concerned, we do not expect that they
render our results -- exact positivity of Wigner and P-function 
after a finite time -- obsolete; because of their exponential smallness
they should not spoil an exact positivity.
Our results should thus be
stable with respect to post-Markovian corrections, but a formal
proof would go beyond the scope of this paper. We hope to return to
these issues elsewhere.


\appendix

\section{Classes of Gaussian pointer states}

We choose the following form for the wave functions 
$\langle q\ket{\G}$ of pure Gaussian 
pointer states $\ket{\G}$:
\be
\psi_{\G}(q)=\left(\a_R/2\pi\right)^{1/4}
  \exp\left(-\a (q-x)^2/4+\I p
   (q-x)\right)\ . \lb{gauss} 
\ee
This is the most general form of a normalized Gaussian wave packet
shifted and boosted uniformly from the fiducial state for $\psi_0(q)$
for $\G=(0,0)$,
\be
\psi_0(q)=\left(\a_R/2\pi\right)^{1/4}\exp(-\a q^2/4)\ . \lb{gauss0} 
\ee
The correlation matrix $\C_{1/4}$, defined in (\ref{C}), takes the
following form:
\be\lb{C1}
\bra{0}
\left(\begin{array}{cc}\hat x^2&(\hat x\hat p+h.c.)/2\\
(\hat x\hat p+h.c.)/2&\hat p^2\end{array}\right)\ket{0}
=\frac{1}{\a_R}\left(\begin{array}{cc}1&-\a_I/2\\
                   -\a_I/2&\vert\a\vert^2/4\end{array}\right)\ .
\ee
Indeed, its determinant is $1/4$ for all $\alpha$.
In \cite{DK} we have found a distinguished value of
the complex parameter $\a$
for the class of robust pointer states, given by
\be\lb{aHS}
\a_0\equiv\a_R+\I\a_I=(1-\I)\sqrt{2Dm}\ .
\ee
For this value (\ref{C1}) gives
\be\lb{C2}
\C_{1/4}=
\left(\begin{array}{cc}\s_0^2/\sqrt{2}&1/2\\
                1/2&\s_0^{-2}/\sqrt{2}\end{array}\right)\ .
\ee
This is being used in the proof for the positivity of
the P-function, Eq.~(\ref{Ppos}). The proof can be performed
for any $\alpha$ (it always leads to a cubic equation),
but the basic object is 
the matrix $\C_{1/4}$ with determinant 1/4.

It is worthwhile to add the following. Let us assume that we had 
constructed no pointer states first. Instead, assume that we had taken an
arbitrary real positive matrix of determinant $1/4$ to
perform a coarse-graining on the Wigner function like in (\ref{QgW}).
This matrix would then define a complex parameter $\a$ as on the
right-hand side of (\ref{C1}), and we 
would be able to construct a unique overcomplete set of
Gaussian wave functions (\ref{gauss}) whose $\C_{1/4}$-matrix is just
our chosen one. Thus it would turn out that our coarse-grained Wigner 
function were just the corresponding Q-function which is always positive.
In this way we have obtained an alternative proof of the theorem
(\ref{theoremW}) of Sect.~2. 

\section{Symplectic Fourier transform}

This appendix is devoted to the proof of (\ref{WgP}) and (\ref{QgW}).
To facilitate the mathematical derivations we use the Fourier 
representation. For instance, the Fourier transformed Wigner function
reads
\be
\tilde W(\Gt)
=\int W(\G)\exp\left[-\I\Gt^T\G\right]\D\G\ , \lb{Fou}
\ee
where $\Gt=(\pt,-\xt)$. We use the symplectic product $\Gt^T\G=\pt x-\xt p$ 
of the original and the Fourier-transformed variables in (\ref{Fou}).
In Fourier representation, the coarse-graining relations 
(\ref{WgP},\ref{QgW}) reduce to algebraic ones:  
\be
\tilde W(\Gt)=\exp\left[-\frac{1}{2}\Gt^T\C_{1/4}\Gt\right]\tilde P(\Gt)\ ,
\lb{WgPFou}
\ee
\be
\tilde Q(\Gt)=\exp\left[-\frac{1}{2}\Gt^T\C_{1/4}\Gt\right]\tilde W(\G)\ .
\lb{QgWFou}
\ee

We are going to prove the first relationship (the second one can easily
be proven along the same lines). From (\ref{Wigner}) and 
(\ref{Fou}) we obtain
\be\lb{WFou}
\tilde W(\Gt)=\int\E^{i\pt x}\bra{x+\half\xt}\r\ket{x-\half\xt}\D x\ .
\ee
Substituting the P-function expansion (\ref{P}) of $\r$
 on the right-hand side yields
\be 
\tilde W(\Gt)=
\int\E^{-i\pt x}                \bra{x+\half\xt}\G^\prime\rangle
                \langle\G^\prime\ket{x-\half\xt}
P(\G^\prime)\D\G^\prime\ .
\ee
Invoking the expression (\ref{gauss}) of the pointer state
wave functions we can perform the above integral, yielding
\be
\tilde W(\Gt)=\exp
\Bigl[-\frac{\a_R\xt^2}{8}
      -\frac{1}{2\a_R}(\pt-\half\a_I\xt)^2\Bigr]\tilde P(\Gt)\ \ .
\ee
The quadratic form in the above exponent can be written as
$-\half\Gt^T\C_{1/4}\Gt$.
 This completes the proof. 

\section{`Fokker-Planck' equation for the P-function}

In this part we return to the notations of our recent work
\cite{DK} and derive directly the evolution equation for the
P-function. 
We introduce the projectors on the states $\psi_{\G}(q)\equiv
\langle q|\G\rangle$, $\P(\G)=|\G\rangle\langle\G|$.
We also denote the P-function $P(\G;t)$ of the open
system by $f(\G;t)$. 
Application of (\ref{master}) to (\ref{P}) yields
\be
\frac{\D}{\D t}\r=\int f(\G;t)\L\P(\G)\D\G
        =\int\dot{f}(\G;t)\P(\G)\D\G\ . \lb{rhodot}
\ee
To derive from this the equation for $f$, we first observe that
the action of $\L$ on $\P(\G)$ in coordinate representation yields
\bea
\bra{q}\L\P\ket{q'} &=& \frac{\I}{2m}
\Biggl\{-\I\a_I-\I p\left[\a^{*}(q'-x)+\a(q-x)\right] \nonumber\\ 
& &\ -\left(\frac{\a^*}{2}\right)^2(q'-x)^2
     +\left(\frac{\a}{2}\right)^2(q-x)^2\Biggr\}\bra{q}\P\ket{q'}
                                                           \nonumber\\
& &\ -\frac{D}{2}(q-q')^2\bra{q}\P\ket{q'}\ . \lb{LP}
\eea

Using~(\ref{gauss}) in the identity
$\bra{q}\P(\G)\ket{q'}=\psi_\G(q)\psi_\G^*(q')$, the above action
can be expressed as an operator
containing derivative terms in $x$ and $p$,
acting on $\P$. After partial integration in (\ref{rhodot})
one then finds the following equation for $f$,
\be
\frac{\D f(\G;t)}{\D t}=-\frac{p}{m}\partial_{x} f(\G;t)
        +\frac{1}{2}\left[ D_{pp}\partial^2_{pp}
                         +D_{xx}\partial^2_{xx}
                        +2D_{px}\partial^2_{px} \right]f(\G;t)
\ , \lb{FP}
\ee
where the elements of the diffusion matrix are given by
\be
{\bf D}\equiv
\left(\begin{array}{cc}D_{xx}&D_{xp}\\D_{px}&D_{pp}\end{array}\right)
=\left(\begin{array}{cc}
-\a_I/m\a_R &\vert\a\vert^2/4m\a_R\\
\vert\a\vert^2/4m\a_R & D\end{array}\right)\ . \lb{D}
\ee
Eq.~(\ref{FP}) can be interpreted as a Fokker-Planck equation
for $f(\G,t)$ provided the
diffusion matrix ${\bf D}$ is non-negative. As usual, the 
first term in (\ref{FP}) will then describe a drift according
to the free-particle dynamics, while the second term
will describe a diffusion of the state of the system over the
pointer states $\P(\G)$. In \cite{DK} we have implemented
the principle of minimal local entropy production by minimising the
width of the Gaussian pointer states. This leads to a value
for $\a$ of the order of (\ref{aHS}). We have shown that the relation
$\C={\bf D}t_0^2/2$ holds for the distinguished value
(\ref{aHS}) of $\a$.

To find a formal solution of~(\ref{FP}), we use the Fourier 
representation $\tilde f(\Gt;t)$.
Eq.~(\ref{FP}) then leads to the following evolution equation
for the Fourier components:
\be
\frac{\D {\tilde f}(\Gt;t)}{\D t}=
 -\frac{\pt}{m}\partial_{\xt}{\tilde f}(\Gt;t)
        -\frac{1}{2}\left[\Gt^T{\bf D}\Gt\right]
        {\tilde f}(\Gt;t)\ . \lb{FPFou}
\ee
The solution assumes the form
\be
{\tilde f}(\Gt;t)
=\exp\left[-\frac{t}{2}\Gt^T{\bf D}(t)\Gt\right]
\tilde f(\xt-\pt t/m,\pt;0)\ . \lb{ftsol}
\ee 
By substitution into (\ref{FPFou}) one obtains explicitly the matrix of 
time-dependent coefficients \cite{Err}
\be
{\bf D}(t)=
\left(\begin{array}{cc}D_{xx}+D_{xp}t/m+D_{pp}t^2/3m^2&D_{xp}+D_{pp}t/2m\\
                       D_{xp}+D_{pp}t/2m&D_{pp}
               \end{array}\right)\ . \lb{Gt}
\ee
This matrix becomes more and more positive for $t>0$ provided  
${\bf D}\equiv{\bf D}(0)$ was chosen positive. One obtains the
solution of the evolution equation (\ref{FP}) 
by the inverse Fourier-transform of the expression (\ref{ftsol}).
It takes the form of the convolution
\be
f(\G;t) = g\left(\G;t{\bf D}(t)\right)*f(x-pt/m,p;0) \ .\lb{fsol}
\ee
The solution therefore emerges as the progressive time-dependent 
Gaussian coarse-graining of the free kinematic evolution. 
Note the close similarity of this equation with the
time evolution (\ref{Wsol}) for the Wigner function.

\section*{Acknowledgements}
We thank Boris Tsirelson for his comments on our manuscript.
L.D. thanks the Hungarian OTKA Grant No. 032640 for financial support.

\end{document}